\documentclass[review]{elsarticle}

\usepackage{lineno,hyperref}
\usepackage{amsmath, amsthm, amssymb}
\usepackage{bbm}
\usepackage{graphicx}
\usepackage{float}
\usepackage{subfig}
\usepackage[margin=1in]{geometry}

\usepackage{textcomp}
\usepackage{epstopdf}

\modulolinenumbers[5]

\makeatletter
\def\ps@pprintTitle{%
  \let\@oddhead\@empty
  \let\@evenhead\@empty
  \def\@oddfoot{\reset@font\hfil\thepage\hfil}
  \let\@evenfoot\@oddfoot
}
\makeatother

\begin{document}

\begin{frontmatter}

\title{Bayesian methods for event analysis of intracellular currents}

\author[aff1,aff2]{Josh Merel\corref{mycorrespondingauthor} \tnoteref{equal}}
\cortext[mycorrespondingauthor]{corresponding author: jsmerel@gmail.com
\\ \textcopyright 2016. This manuscript version is made available under the CC-BY-NC-ND 4.0 license \url{http://creativecommons.org/licenses/by-nc-nd/4.0/}
\\ accepted for publication in Journal of Neuroscience Methods, DOI: 10.1016/j.jneumeth.2016.05.015
}
%\\ \indent website: https://sites.google.com/site/jsmerel/
\tnotetext[equal]{These authors contributed equally.}
\author[aff3]{Ben Shababo \tnoteref{equal}}
\author[aff3]{Alex Naka}
\author[aff3,aff4]{Hillel Adesnik}
\author[aff1,aff2,aff5,aff6]{Liam Paninski}
%\address{Room 1005 SSW, MC 4690, New York, NY 10027} %stat department address

%fnref and fntext
%corref and cortext
%\tnoteref and \tnotetext
%\author[mymainaddress,mysecondaryaddress]{Center for Theoretical Neuroscience and Department of Statistics
%Columbia University Neurobiology and Behavior program, Columbia University }
\address[aff1]{Neurobiology and Behavior program, Columbia University}
\address[aff2]{Center for Theoretical Neuroscience, Columbia University}
\address[aff3]{Helen Wills Neuroscience Institute, University of California, Berkeley}
\address[aff4]{Department of Molecular and Cellular Biology, University of California, Berkeley}
\address[aff5]{Department of Statistics, Columbia University}
\address[aff6]{Grossman Center for the Statistics of Mind, Columbia University}


% Highlights:
%We present a Bayesian approach for automatic event analysis.
%The method was designed and validated for voltage-clamp recordings. 
%The method outperforms existing methods on simulated and real data.
%We demonstrate extensions useful for synaptic mapping experiments.

\begin{abstract}
\textbf{Background:}
Investigation of neural circuit functioning often requires statistical interpretation of events in subthreshold electrophysiological recordings. This problem is non-trivial because recordings may have moderate levels of structured noise and events may have distinct kinetics.  In addition, novel experimental designs that combine optical and electrophysiological methods will depend upon statistical tools that combine multimodal data.
\\ \textbf{New method:}
We present a Bayesian approach for inferring the timing, strength, and kinetics of post-synaptic currents (PSCs) from voltage-clamp electrophysiological recordings on a per event basis.  The simple generative model for a single voltage-clamp recording flexibly extends to include additional structure to enable experiments designed to probe synaptic connectivity. 
\\ \textbf{Results:}
We validate the approach on simulated and real data.  We also demonstrate that extensions of the basic PSC detection algorithm can handle recordings contaminated with optically evoked currents, and we simulate a scenario in which calcium imaging observations, available for a subset of neurons, can be fused with electrophysiological data to achieve higher temporal resolution.
\\ \textbf{Comparison with existing methods:}
We apply this approach to simulated and real ground truth data to demonstrate its higher sensitivity in detecting small signal-to-noise events and its increased robustness to noise compared to standard methods for detecting PSCs.  
\\ \textbf{Conclusions:}
The new Bayesian event analysis approach for electrophysiological recordings should allow for better estimation of physiological parameters under more variable conditions and help support new experimental designs for circuit mapping.

\end{abstract}


\end{frontmatter}

%\linenumbers

\section{Introduction}

Subthreshold neuronal activity provides an unsurpassed richness of information about a single cell's physiological properities. Access to subthreshold activity allows for inference about intrinsic biophysical properties (e.g. membrane and ion channel parameters), circuit level properties (e.g. synaptic connectivity), neural coding (e.g. receptive fields), and synaptic properties (e.g. quantal properties \& plasticity).  At present, whole-cell patch-clamp stands alone in its ability to reliably access subthreshold activity owing to excellent signal-to-noise ratio (SNR) and very high temporal precision, as opposed to optical subthreshold measurements.  At the same time, optical technologies have advanced to the point where we can observe the suprathreshold activity of hundreds of individual neurons simultaneously with calcium imaging and stimulate neurons by subtype or spatial location \cite{Rickgauer2014}.  However, the limits on temporal resolution and the indirectness of the observations make inferring fine-scale network and cellular parameters difficult. Approaches which combine optical tools with electrophysiology offer unique advantages \cite{scanziani2009electrophysiology}.    
In this work we present new statistical techniques useful for analyzing whole-cell data as well as extensions demonstrating how our approach is particularly well-suited to settings where electrophysiology is combined with optical physiology.

\subsection{Our setting and approach}
Fundamentally, many of the subthreshold-based analyses mentioned above depend on the interpretation of the recorded time series as a sequence of \textit{events}. 
In this setting, events are the successful transmission of neurotransmitter onto the recorded cell, and when this occurs, a transient current flows into or out of the cell, known as a postsynaptic current (PSC).  The analyses of experiments designed to infer properties of evoked or spontaneous inputs to a cell (e.g. monosynaptic mapping or quantal/mini-PSC analyses) require determining when a postsynaptic event happened and describing that event. 
Estimating PSC properties is most straightforward when recordings are acquired using the voltage-clamp configuration which employs a feedback circuit to hold the membrane potential at a constant value thus mitigating variability in PSC properties due to the intrinsic biophysics of the cell (though see \cite{baryehuda1127}).

In this work, we present a Bayesian approach for inferring the timing, strength, and kinetics of postsynaptic currents from voltage-clamp recordings, and we demonstrate on simulated and real data that this method performs better than standard methods for detecting PSCs.
The improvement in single-trial accuracy with our method should allow for better estimation of physiological parameters with less data and under more variable conditions (e.g. when the exact timings of stimuli or its effects are unknown).
In addition, the quantification of uncertainty over PSC features provided by Bayesian inference enables new experimental designs (e.g. \cite{shababo2013bayesian}).

Bayesian approaches are naturally extensible, so the intuitive, generative model and straightforward inference procedure flexibly extend to include structure relevant to the analyses mentioned above. Specifically, we extend the core single-trial model to include types of data obtained in monosynaptic mapping experiments which may involve optical stimulation artifacts or combine voltage-clamp recordings and optical recordings. For this latter extension, we combine the single trace model presented in this work with related work on calcium imaging \cite{pnevmatikakis2013bayesian} to demonstrate a Bayesian approach to analyzing mapping experiments consisting of simultaneous population calcium imaging and single cell voltage-clamp recordings \cite{aaron2006reverse}.

\subsection{Review of other approaches}

To our knowledge, all previous methods for inferring PSCs have relied on first inferring the timing of single events (i.e. event onsets), and then sometimes fitting per-event kinetics given that event time. These methods have tended to fall into two categories.  The superficially simpler of the two approaches is find events by thresholding the trace or its first derivative (i.e. finite difference).  In practice, such methods have extra parameters for smoothing, computing the appropriate offset, or post-processing.  Implementations tend to over-detect candidate events and then evaluate candidates based on analysis of per event kinetics \cite{jonas1993quantal,ankri1994automatic,hwang1999automatic,kudoh2002simple}.  For concreteness, consider a two-stage approach wherein a threshold is used to identify initial candidates, and then a model is fit to the transient dynamics in order to confirm or reject candidate events by comparison of the parameters of the dynamics against pre-determined criteria \cite{ankri1994automatic}.  Even with post-processing, such methods can be non-selective and tend not to exploit all of the available information (i.e. the transient dynamics aren't used to detect the events initially).  

Threshold methods have been largely superseded by the second class of approaches, template-based methods  \cite{clements1997detection,pernia2012deconvolution}. In these methods, templates are usually learned by averaging event-responses collected by a simpler method (e.g. thresholding and/or hand-curation).
While template methods are straightforward, initial attempts to apply these methods failed when the amplitude of the events varied or where events overlapped - both common scenarios.  
The first commonly used algorithm for PSC detection that attempted to avoid issues related to amplitude variability introduced the idea of rescaling a fixed template at each time step \cite{clements1997detection}.
Following this trend, template-matching approaches have gradually shifted towards deconvolution methods, which are a more well-founded way to use templates \cite{pernia2012deconvolution}. Deconvolution generally refers to methods that assume the observed trace is the result of convolving a template with unobserved events (of varying amplitude), and such methods invert this model to estimate the times from the template.
Both of these template-based methods produce inferred events with different amplitudes and a threshold can then be used to screen out small events  (see \cite{guzman2014stimfit} for a Python implementation of \cite{clements1997detection} and \cite{pernia2012deconvolution}, and see \cite{richardson2008measurement} for deconvolution of current clamp traces).

Methods that rely on fixed-shape templates can work very well when the shape of the event is consistent across events, but postsynaptic events can vary in shape and amplitude, especially for events from different pre-synaptic sources due to different dendritic filtering, issues with space-clamp, or different receptor subunit distributions.  Indeed, a core rationale behind the initial preference for threshold based approaches was the recognition that events may vary too much for a single template.
While it is possible to use approaches that employ multiple templates \cite{li2007weighted,shi2010novel}, there are still potential issues related to the stage-wise separation between learning the template and subsequent detection causing a sub-optimal use of information. 

We take a Bayesian approach, rooted in a probabilistic, generative model. Broadening the taxonomy, this approach is a type of deconvolution method.  However, we do not consider a single template (or a handful of templates), but instead a distribution over templates through the use of prior distributions on the kinetics and amplitudes of individual PSCs. Importantly, we also model event timing in continuous time (i.e. without binning), and we incorporate an autocorrelated, AR($p$) noise process \cite{chib1994bayes}, which provides a more accurate description of the data. This leads to more precise detection of event times and inference that is more robust (i.e. less susceptible to noise).  As such, our inference better leverages all available information (i.e. all events and full timecourse of each event).  Given this probabilistic formulation of the noise process and the inclusion of priors on the PSC features, we can then perform posterior inference in this model using Markov chain Monte Carlo (MCMC, see methods).

A tradeoff is that the proposed approach is more computationally intensive than previous approaches. Nevertheless, we believe the flexibility and robustness that this approach affords makes up for this in many settings.  Beyond handling overlapping events and variation in the shape of events, our method inherits advantages of probabilistic modelling. The method is extensible and amenable to serving as a modular component of hierarchical models, as we show.  
Moreover, while existing methods tend to produce all-or-none results and the precise timing of the event is a secondary consideration, using a probabilistic approach, it is straightforward to consider posterior uncertainty. We essentially get a level of confidence for detection of each event and the level of uncertainty in the precise timing of the event.  This posterior uncertainty in event times can translate into posterior uncertainty for other parameters of interest, such as synaptic weights. 

\subsection{Overview}
In the following sections, we will first present the details of the model for single-trial voltage-clamp traces and extensions mentioned previously.  We then provide the details of the inference scheme we use for sampling from the posterior distribution.  
{In the results, we compare our approach with the standard template-based approach \cite{clements1997detection} as well as a Wiener Filtering approach \cite{wiener1949}, since these serve as competitive and robust baselines.  
Note that while we are not aware of the Wiener filter having been explicitly proposed for this application, the Wiener filter is a standard deconvolution approach (similar to \cite{pernia2012deconvolution}).  Unlike the deconvolution approach in \cite{pernia2012deconvolution}, the Wiener filter automatically and optimally determines deconvolution parameters from the power spectral density (PSD), and we found this performs better over a range of data than a deconvolution approach with hand-tuned parameters (not shown).}  
We show inference results from our approach on simulated and real data for spontaneous EPSCs and IPSCs across several cell types, PSCs evoked via paired-patching as ground-truth validation, and PSCs evoked optically with one-photon and two-photon stimulation with stimulation artifacts. We also show results on simulated data for a mapping experiment which combines voltage-clamp recordings with calcium imaging, illustrating extensibility.

\section{Methods}

We draw on tools developed in statistics \cite{moller2004} and signal processing \cite{tan2008} to decompose a voltage-clamp recording into interpretable elements. 
In this application, our events are unitary synaptic currents and their features describe the strength and kinetics of each event. In previous work, we have found similar methods useful for inferring spiking events in calcium imaging data \cite{pnevmatikakis2013bayesian}.

The framework involves (1) specifying a generative model for voltage-clamp recordings, the parameters of which describe event times, features, the noise model, etc., and (2) performing Bayesian inference on event times and features and the model parameters jointly.  Theoretically, Bayesian estimators have nice guarantees (under a ``true" model, see \cite{lehmann1998theory}).  
{In practice, the generic Bayesian approach can fail if the model is inadequate (e.g. in our case, if the generative model does not capture the true statistics of voltage-clamp traces) or if the inference algorithm performs poorly and the true model posterior is not obtained. Motivated by these legitimate concerns, it is critical to validate that for our application, the model captures the statistics of real voltage-clamp data and that inference performs well on both simulated and real data (see Results).}  

\subsection{Model of a single electrophysiological trace} 

In the simplest version of our model, the observed current trace, $y_t$, is a discrete time series composed of the sum of $n$ unitary synaptic currents, a baseline (holding current), $b$, and observation noise $\epsilon_t$ (eq. \ref{sum_of_events}).

\begin{align}
y_{t} &= \sum_{i=1}^n a_{i} f_i(t-t_{i}) + b + \epsilon_t. \label{sum_of_events} \\ 
f_i(t) &= (e^{-t/\tau_i^{d}} - e^{-t/\tau_i^{r}})\mathbbm{1}(t>=0). \label{template_eqn} \\
\epsilon_t &= \sum_{j=1}^p \phi_j \epsilon_{t-j} + u_t, \hspace*{.5in} u_t \sim \mathcal{N}(0,\sigma^2). \label{AR_noise}
\end{align}

\begin{figure}[t]
\begin{center}
  \subfloat{\includegraphics[width=1\textwidth]{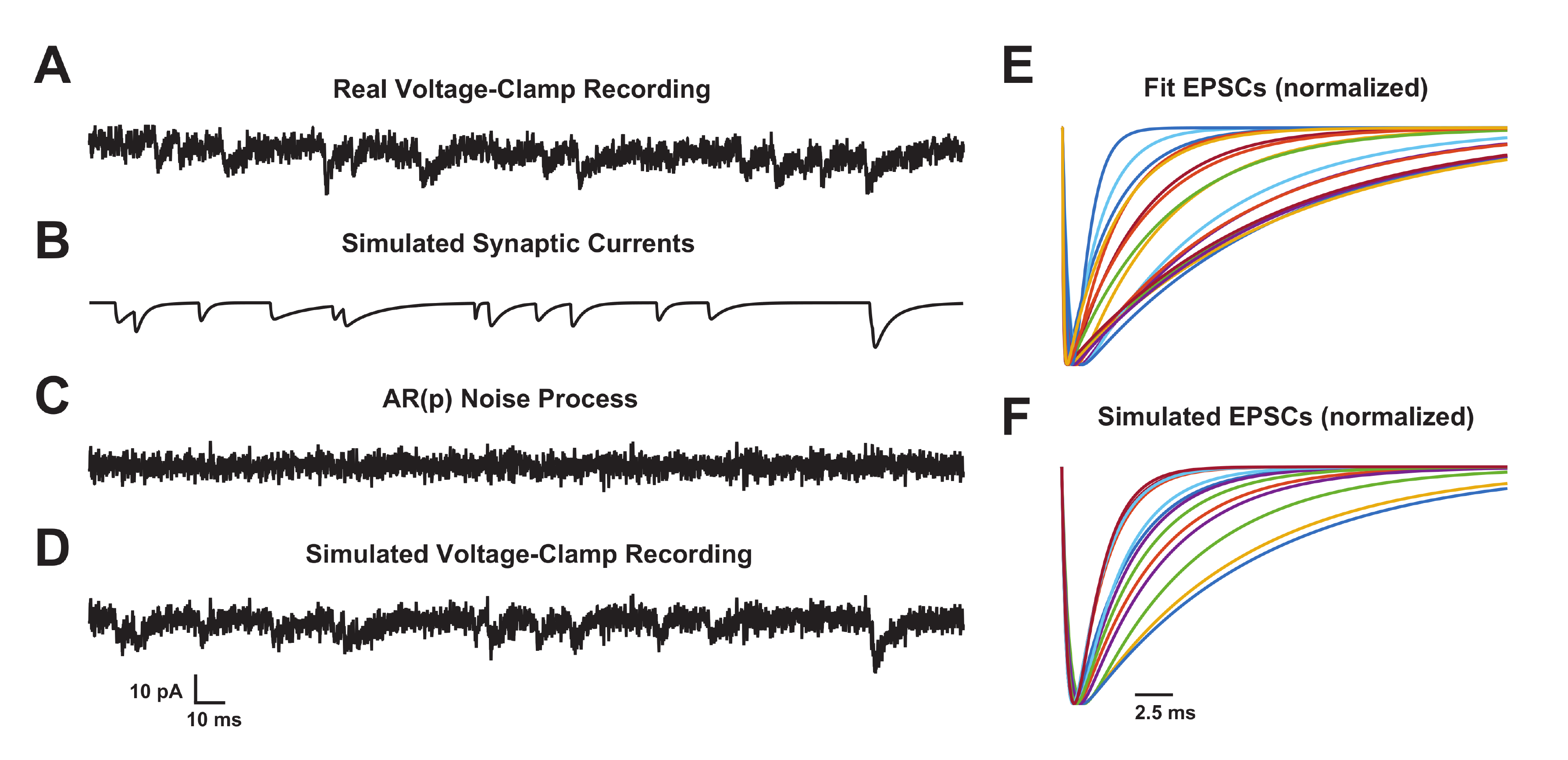}}  
  \end{center}    
  \captionsetup{width=1\textwidth}
\captionsetup{font={footnotesize}}
\caption{This figure depicts the correspondence between real data and the generative model.  \textbf{A} Shows a real voltage-clamp recording. \textbf{B} Depicts simulated synaptic currents generated from a fit to the data in \textbf{A} (noiseless). \textbf{C} Depicts AR($p$) noise process ($p=2$) generated from a fit to the data in \textbf{A}. \textbf{D} Illustrates a model-based voltage-clamp recording simulation (the sum of the data in \textbf{B} and \textbf{C}), to illustrate that the simulation visually captures core features of the real data. \textbf{E} Shows individual events estimated from the data in \textbf{A} (with normalized amplitudes), and panel \textbf{F} shows the individual simulated events used in \textbf{B} (with normalized amplitudes).}
\label{fig:model_demo}
\end{figure}

Each event, indexed by $i$, is characterized by an event time, $t_i \in \mathbb{R}^{+}$, which need not be aligned with the sampling time of $y_t$, its own kinetics determined by $f_i(\cdot)$, and a strength which we define as the amplitude, or peak current, of the event, $a_{i}$.  For synaptic currents, we use a difference of exponentials for $f_i$ which is parameterized by a rise time constant, $\tau_i^{r}$, and a decay time constant, $\tau_i^{d}$. 
For an example of the model, see Figure \ref{fig:model_demo}\textbf{B}-\textbf{D}.  Observe in Figure \ref{fig:model_demo}\textbf{E} that recovered kinetics of individual events do vary significantly, suggesting that a model which captures this structure should perform better than methods which rely on a single (or handful of) template(s).

As opposed to an \textit{i.i.d.} Gaussian noise process, the more general autoregressive, AR($p$), process better captures the noise in voltage-clamp recordings.  We have found the noise model to be crucial for robust inference (see Results). 
In an AR($p$) noise model, the noise has temporal correlations due to direct dependencies between noise values for $p$ timesteps. In this work, we use an AR noise model with $p = 2$ (eq. \ref{AR_noise}). Voltage-clamp recordings exhibit correlated noise whose source can be electrical hardware, changes in resistance between the electrode and the interior of the neuron, and other biological non-event contributions to the observation.
In practice, it is these forms of temporally correlated noise which lead to many of the false positives since they are more likely to exhibit a similar shape to true events. 

With eqs. \ref{sum_of_events} and \ref{AR_noise} we can write down the likelihood of the observed, noisy data given the parameters 
{($\Theta \equiv \{\sigma,b,n,\{a_i,t_i,\tau_i^{d}, \tau_i^{r}\}_{i=1...n}, \{\phi_j\}_{j=1...p}\}$)}. We begin with the i.i.d, i.e. AR($0$), case,
\begin{equation}
p(Y|\Theta) = \prod_{t=1}^T (2\pi\sigma^2)^{-1/2} exp[-\frac{1}{2\sigma^2}(y_t - \hat y_t)^2],
\end{equation}
where $\hat y_t$ refer to the predicted noiseless trace:
\begin{equation}
\label{ar0_likelihood}
\hat y_t = \sum_{i=1}^n a_{i} f_i(t-t_{i}) + b.
\end{equation}
In equation 8 of \cite{chib1994bayes}, the likelihood is extended to the AR($p$) case
\begin{equation}
\label{main_likelihood}
p(Y|\Theta) = \prod_{t=1}^T (2\pi\sigma^2)^{-1/2} exp[-\frac{1}{2\sigma^2}(y_t - \hat y_{t|t-1})^2],
\end{equation}
where $\hat y_{t|t-1}$ is (adapted from equation 11 of \cite{chib1994bayes}, {omitting boundary observations}),
\begin{equation}
\hat y_{t|t-1} = \hat y_t + \sum_{j=1}^p \phi_j (y_{t-j} - \hat y_{t-j}).
\end{equation} 

The probabilistic model provides a natural objective function:  
\begin{equation}
\mathcal{L}(\Theta | Y) \propto \ln p(Y|\Theta) + \ln p(\Theta).
\end{equation} 

It is possible to optimize this log-posterior directly, or inference can be performed to obtain an estimate of the posterior distribution. $p(\Theta)$ corresponds to the prior probability on the parameters. 
{In a probabilistic formulation, it is worth explicitly keeping in mind that the posterior distribution for a given parameter can only have support where its prior distribution has support, so hard constraints (e.g. a parameter being positive or a minimum amplitude size) are naturally incorporated as prior information. 
For amplitudes, baselines, time-constants, and event times, we use non-informative, improper uniform priors over either real or bounded real numbers (we selected very broad ranges appropriate for each parameter).  
The prior on the event count is given by a Poisson distribution which has one free parameter corresponding to the prior expected number of events -- this is a parameter that a user would tune depending on their general expectation about the number of events in their data.
The prior on the noise level $\sigma^2$ is a diffuse inverse gamma distribution (which is the conjugate prior), and we use a diffuse prior distribution over $\phi$ (i.e. a normal distribution subject to stability constraints, same as \cite{chib1994bayes}) -- see section \ref{section:Inference} on inference for more details.  

Additionally, we note that more sophisticated priors in a Bayesian model serve as natural ways to extend the core model.}
For example, distributions over event features could be modulated by clustering onto presynaptic sources or the rate of events in each trace could be time-varied based on stimulation of presynaptic cells in mapping experiments.  {Model extensions are explored in the next section.}

\subsection{Model extensions}

\subsubsection{Including optical currents}

In this extension, we consider an active mapping experiment where we have some level of spatially structured optical stimulation (via optogenetics \cite{fenno2011} or neurotransmitter uncaging \cite{callaway2002}) of presynaptic cells while holding a postsynaptic cell in voltage-clamp \cite{Shepherd2005,Katzel2011}. In this setting, detections of PSCs coincident with stimulations can be used to infer connectivity between neurons. However, in some protocols, the postsynaptic cell also responds to the stimulation\footnote{This is seen with glutamate uncaging at locations near the patched cell and could also occur with optogenetics when mapping connections between cells of the same transcriptional identity in the same location or when mapping many heterogeneous populations of cells (i.e. when a pan-neuronal promoter is be used).}.
If this is the case, we will see a direct optically evoked current in our voltage-clamp recording when we attempt to stimulate cells near the patched cell. In order to remove artifacts of this sort, we can incorporate a parameterized, additive term in the generative model and then perform inference jointly with respect to these parameters. The choice of an additive term is more reasonable for optogenetic stimulation because the currents are carried through different channels whereas with neurotransmitter uncaging the direct stimulation current and the synaptic current may be competing for the same channels.

For example, it is straightforward to include a parameterized kernel for the optical response of the neuron, $h(\cdot)$, and then to convolve that response with the known optical input to the neuron (e.g. the laser or LED power),
\begin{align}
y_{t} &= \sum_{i=1}^n a_{i} f_i(t-t_{i}) + \sum_{j}^{n_s} a_j h(\theta_h)\ast d_j(t) + b + \epsilon_t,
\label{artifact_w_shape_eqn}
\end{align}
where we have $n_s$ optical inputs each with known timecourse $d_j(t)$ and the response kinetics are modelled with a convolution which is parameterized by $\theta_h$ (e.g. a set of time constants). Each stimulation will have its own gain, $a_s$, which depends on the density of the corresponding channels at the location of that stimulation. We have found this approach to be useful under certain conditions with optogenetics (Figure \ref{fig:direct_stim_real}\textbf{D}).
However, for some optogenetic currents, the multi-state kinetics of opsins can make it difficult to design a parameterized $h(\cdot)$ which can be sampled efficiently. 
In this case, the shape of the optical current could be measured empirically and then modelled as
\begin{align}
y_{t} &= \sum_{i=1}^n a_{i} f_i(t-t_{i}) + \sum_{j=1}^{n_s} a_j h(t-t^{(s)}_{j}) + b + \epsilon_t,
\label{artifact_eqn}
\end{align}
where $h(\cdot)$ is now the stereotyped shape of the current and we know the set of times $\{t^{(s)}_j\}_{j=1...n_s}$ at which we have stimulated. Since the currents are filtered in the dendrites, eq. \ref{artifact_eqn} breaks down slightly when handling stimulations at locations at varying distances from the soma of the patched cell (Figure \ref{fig:direct_stim_real}\textbf{D}). Nonetheless, it still provides a good trade off between accuracy and computational tractability. In addition, one could create several optical current templates based on the distance of the stimulation site from the soma \cite{mena2015}.

\subsubsection{Mapping with calcium imaging and voltage-clamp recordings} 

Next we extend the model to show how the inferred events could be identified with presynaptic sources when the local population is partially observed via calcium imaging \cite{aaron2006reverse} (alternatively, a similar approach could combine information from voltage imaging).
In this setting our observations will consist of the fluorescence traces of the imaged cells and the voltage-clamp recording. Both the presynaptic fluorescence traces, $c_t^k$ where $k$ indexes the neurons observed via imaging, and the postsynaptic electrophysiological recording, $y_t$, can be interpreted as a sum of events \cite{pnevmatikakis2013bayesian},
  
\begin{align}
c_{t}^{k} &= \sum_{i=1}^{n_k} a^{(c)}_{k} g_k(t-t_{ki}) + b_k + \eta^k_t, \\
y_{t} &= \sum_{k=1}^K \sum_{i=1}^{n_k} a^{(y)}_{k} f_k(t-t_{ki}) + \sum_{j=1}^{n_{\emptyset}} a^{(\emptyset)}_{j} f_j(t-t_{j}) + b + \epsilon_t,
\label{mapping_model}
\end{align}

\noindent with $\epsilon_t$ as before and $\eta^k_t \sim \mathcal{N}(0,\nu_k^2)$.  Like $f_k(\cdot)$, $g_k(\cdot)$ is also a sum of exponentials but the timescale of the kinetics is much slower.  $\{t_{j}\}_{j=1...n_{\emptyset}}$ are the times of PSCs with no corresponding imaged, presynaptic cell. We want to point out that because we model the process in continuous time, the observed calcium traces need not have the same sampling rate or observation times as each other or the voltage-clamp recording (which will be sampled many orders of magnitude faster).

In this demonstration, we have chosen to assume the calcium events for the same cell are all of equal amplitude -- this would be appropriate if the spikes occur sparsely or if calcium transients sum roughly linearly.  If the summation is known to be nonlinear with some biophysically plausible nonlinearity (e.g. see \cite{vogelstein2009spike}), this could be straightforwardly accommodated by modifying the model and similar inference methods may still be applied.
Alternatively, it is also straightforward for the calcium event amplitudes to vary across events.  

We have similarly chosen for all postsynaptic events to be of the same amplitude and shape when they follow from a specific presynaptic cell, but in practice one can extend the clustering to impose cell-specific priors on these amplitudes and shapes.  Some subset of events observed in the electrophysiological recording will arise from unobserved presynaptic inputs so we allow these to be explained by events with no observed presynaptic cause and with independent shape and amplitude per event (second summation term indexed $i=1...n_{\emptyset}$).  We could also incorporate a fixed delay between presynaptic events and postsynaptic events (this would be an additional parameter, pre-specified or inferred, in the postsynaptic transient response $f_k(\cdot)$).  

Since we now observe multiple traces, our new objective function is simply the sum of the log probabilities of the various traces.  The multiple traces are conditionally independent given event times $\{t_{ki}\}_{k=1...K,i=1...n_k}$ and $\{t_{j}\}_{j=1...n_{\emptyset}}$, so we have an overall objective corresponding to the log-posterior:
\begin{equation}
\mathcal{L}(\Theta | C^k,Y) \propto \ln p(Y|\Theta) + \sum_{k=1}^K  \ln p(C^k|\Theta) + \ln p(\Theta),
\end{equation} 

\noindent again with $p(\Theta)$ corresponding to the prior probability on the full set of parameters (see Section \ref{section_passive_mapping} for results).

\subsection{Inference}
\label{section:Inference}

For this work, we perform inference using Markov chain Monte Carlo (MCMC) \cite{neal1993probabilistic,gelman2014bayesian}. 
MCMC techniques allow us to obtain samples from the posterior distribution over all unknown variables in the model and thereby approximate the posterior by a histogram of such samples.
Specifically, we perform Gibbs sampling over all the parameters \cite{gelman2014bayesian}.  This means that for each parameter, we hold all other parameters fixed and conditionally update the focal parameter by sampling it from its conditional distribution.  A ``sweep" consists of an update of all parameters.  While it is possible to compute conditional distributions analytically for sufficiently simple models, it is quite simple to use generic sampling methods to update parameters for any model for which one can compute the likelihood.  For example, we use random-walk Metropolis (RWM) to update many parameters -- this consists of updating parameters by proposing updates from a distribution centered on the current value and accepting or rejecting proposed updates such that the resulting set of samples are consistent with the conditional distribution \cite{gelman2014bayesian}.  Alternatively, more powerful samplers such as Hamiltonian Monte Carlo could be used \cite{gelman2014bayesian} but simple RWM sufficed here.
  
In the single-trial, voltage-clamp case, we update $\{t_i,a_i,b,\tau_i^{d}, \tau_i^{r}\}_{i=1...n}$ by RWM.  Inclusion of a direct optical current, in the simplest case, only contributes one additional amplitude parameter for each stimulation, and these can be inferred similarly.  {Through sampling, we also estimate the posterior distribution of the number of events in each trace, given a Poisson prior (note that this could be generalized to an inhomogenous Poisson process prior so that the event rate can vary based on inputs such as optical stimulation)}. To add and remove events (i.e. perform inference over $n$), we use birth-death moves which consist of the proposal of a new event time and the removal of an existing event time respectively \cite{moller2004}.  

The noise process parameters $\phi_{1..p}$ are sampled by rejection sampling from the constrained conditional distribution and $\sigma$ is sampled from its conditional distribution.
Specifically, we reproduce the updates for the $\phi_{1..p}$ and $\sigma$, which are provided in section 4.1 of \cite{chib1994bayes}. $\phi_{1..p}$ is shown to be conditionally normal with a mean and posterior that depend on $\hat e_t = y_t - \hat y_t$ and $\sigma^2$, but also with the constraint that the AR($p$) process defined by $\phi_{1..p}$ is stable {(note that stability here means that the AR process remains bounded on bounded intervals if run autonomously, which can be quickly checked by examining the magnitude of the roots of the associated characteristic polynomial \cite{Lay2012}).  Omitting boundary observations}:
\begin{align}
E &= [\hat e_{t-1} ... \hat e_{t-p}], \\
\Phi_n &= \Phi_0 + \sigma^{-2}(E'E), \\
\hat \phi &= \Phi_n^{-1} (\Phi_0 \phi_0 + \sigma^{-2} E'e), \\
\phi &\sim \mathcal{N}(\hat \phi,\Phi_n^{-1})\mathbbm{1}_{S \phi},
\label{phi_update}
\end{align}
where $\mathbbm{1}_{S \phi}$ is an indicator over the set of stable $\phi$ values and $\phi_0$ and $\Phi_0$ are set to weakly informative prior values.
Following \cite{chib1994bayes}, we place an inverse-gamma prior on $\sigma^2$, which is the conjugate prior given the likelihood (eq. \ref{main_likelihood}) so that we can sample directly from the conditional posterior,
\begin{align}
\delta_1 &= \sum_{t=1}^T (y_t - \hat y_{t|t-1})^2, \\
\sigma^2 &\sim \mathcal{IG}(\frac{1}{2} (T + \nu_0),1/(\frac{1}{2} (\delta_1 + \delta_0))),
\label{sig_update}
\end{align}
where $\nu_0$ and $\delta_0$ are also set to weakly informative prior values.

For the full mapping case, inference is similar.  However, additional moves need to be incorporated to improve mixing.  That is, in addition to the single-variable updates, additional custom moves are performed each sweep.  The additional moves correspond to proposing a swap for the presynaptic identity for a PSC event. That is, an MCMC step is proposed wherein an event associated with one presynaptic source is eliminated and an event at the same time is considered for another source.  To implement such moves, we simply propose to drop an event at a given time for one presynaptic source and to add an event at the same time for another presynaptic source and evaluate the combined acceptance or rejection of these proposed moves using the Metropolis-Hastings ratio \cite{gelman2014bayesian}.    

\subsection{Specific implementation details}

Code implementing our inference routine { is available on github\footnote{\url{https://github.com/jsmerel/joint_calcium_ephys_mapping}}}.  {When applying our inference method to data, we can run from a cold start, or we can initialize event times with those found via a simpler, faster method (e.g. \cite{clements1997detection} or \cite{pernia2012deconvolution})}.  In this work, to establish initial performance of our algorithm, we present results on simulated and real data using cold starts.  
{ MCMC methods ideally will be run long enough that a proper posterior distribution can be estimated.  A standard convergence check for MCMC methods is to run multiple chains and assess that they provide similar estimates.  We did not run multiple Markov chains in this case for all analyses.  Rather, for ease of use, we followed a simple heuristic of initially tuning the total number of sweeps by running multiple Markov chains and then consistently performing a conservatively large number of sweeps to robustly achieve good performance (i.e. as measured relative to other methods).    

For applications, it is pragmatic to adjust the total number of samples to trade off reliability (i.e. high probability of convergence) against computation time. 
For small timeseries (roughly 1 second), inference consisting of 2000 sweeps of the sampler tends to provide an accurate posterior (see Results, simulated data) and requires anywhere from a few seconds to a few minutes on a contemporary computer depending on the number of inferred events}.  
{It is standard to discard the first fraction of the sweeps as \textit{burn-in} (we discard between one-fifth to two-fifths of the total number of sweeps, depending on the event rate and its prior).}
For longer timeseries or many timeseries, we run smaller sections of the timeseries in parallel on a computing cluster {(on a single CPU, the computation time would scale linearly with the number of requested samples and scale linearly in the number of inferred events, which is assumed to be proportional to the length of the trace)}.  We also note that RWM requires a proposal width hyperparameter -- this can be automatically tuned early in the sampling process by adjusting the proposal variance such that the accept rate is reasonable \cite{rosenthal2011optimal}. { We perform this automatic tuning in an \textit{ad hoc} adaptive fashion, incrementally increasing or decreasing the proposal variance when either too many or too few moves are being accepted.}

We also note that in the implementation, we frequently rely on the log-likelihood, $\ln(p(Y|\Theta))$, to determine whether to accept or reject moves.  In order to minimize redundant operations, it proves useful to store $\hat e_t = y_t - \hat y_t$ and we perform evaluations of $y_t - \hat y_{t|t-1}$ by observing that:
\begin{align}
y_t - \hat y_{t|t-1} &= y_t - (\hat y_t + \sum_{j=1}^p \phi_j (y_{t-j} - \hat y_{t-j})) \\
&= y_t - \hat y_t - \sum_{j=1}^p \phi_j (y_{t-j} - \hat y_{t-j}) \\
&= \hat e_t - \sum_{j=1}^p \phi_j \hat e_{t-j}.
\end{align}
{Finally, note that key hyperparameter settings are presented in Supplementary Table 1.}

\subsection{Experimental methods}

All experiments were performed in accordance with the guidelines and regulations of the Animal Care and Use Committee of the University of California, Berkeley. Mice used for experiments in this paper were either wild type (ICR white strain, Charles River), som-IRES-Cre (JAX stock \#018973); Ai9 Rosa-LSL-tdTomato (JAX stock \#007909), PV-Cre (JAX stock \#008069); Ai9 Rosa-LSL-tdTomato, or emx1-IRES-Cre (JAX stock \#005628).
 
\textit{Viral Infection:} Neonatal emx1-cre mice were injected with AAV9-CAG-flexed-Chr2-tdTomato (for 1-photon experiments) or AAV9-syn-ChrimsonR-tdTomato (for 2-photon experiments) at P0-P4. Viruses were acquired from the University of Pennsylvania Vector Core. Undiluted viral aliquots were loaded into a Drummond Nanoject injector. Neonates were briefly cryo-anesthetized and placed in a head mold. With respect to the lambda suture coordinates for S1 were: 2.0 mm AP; 3.0 mm L; 0.3 mm DV. 

\textit{2-Photon Optogenetic Stimulation:} 1040 nm light (femtoTrain, Spectra-Physics) was delivered to the sample using a VIVO 2-Photon workstation (3i) based on a Sutter Moveable Objective Microscope (Sutter, Novato, CA) and the hologram was created using a Phasor 2-Photon computer-generated holography system (3i). Light was delivered for 10 milliseconds at 100 mW power on sample. The hologram for these data was a disc of radius 15 $\mu$m.

For more detailed methods on brain slicing, \textit{in vitro} and 
\textit{in vivo} electrophysiology, and 1-photon optogenetic stimulation, see \cite{Pluta2015}.
{All electrophysiology was analyzed at 20 kHz, or equivalently, with timebins of 0.05 ms.}

\section{Results}

\subsection{AR noise model validation}

\begin{figure}[!h]
\begin{center}
  \subfloat{\includegraphics[width=1\textwidth]{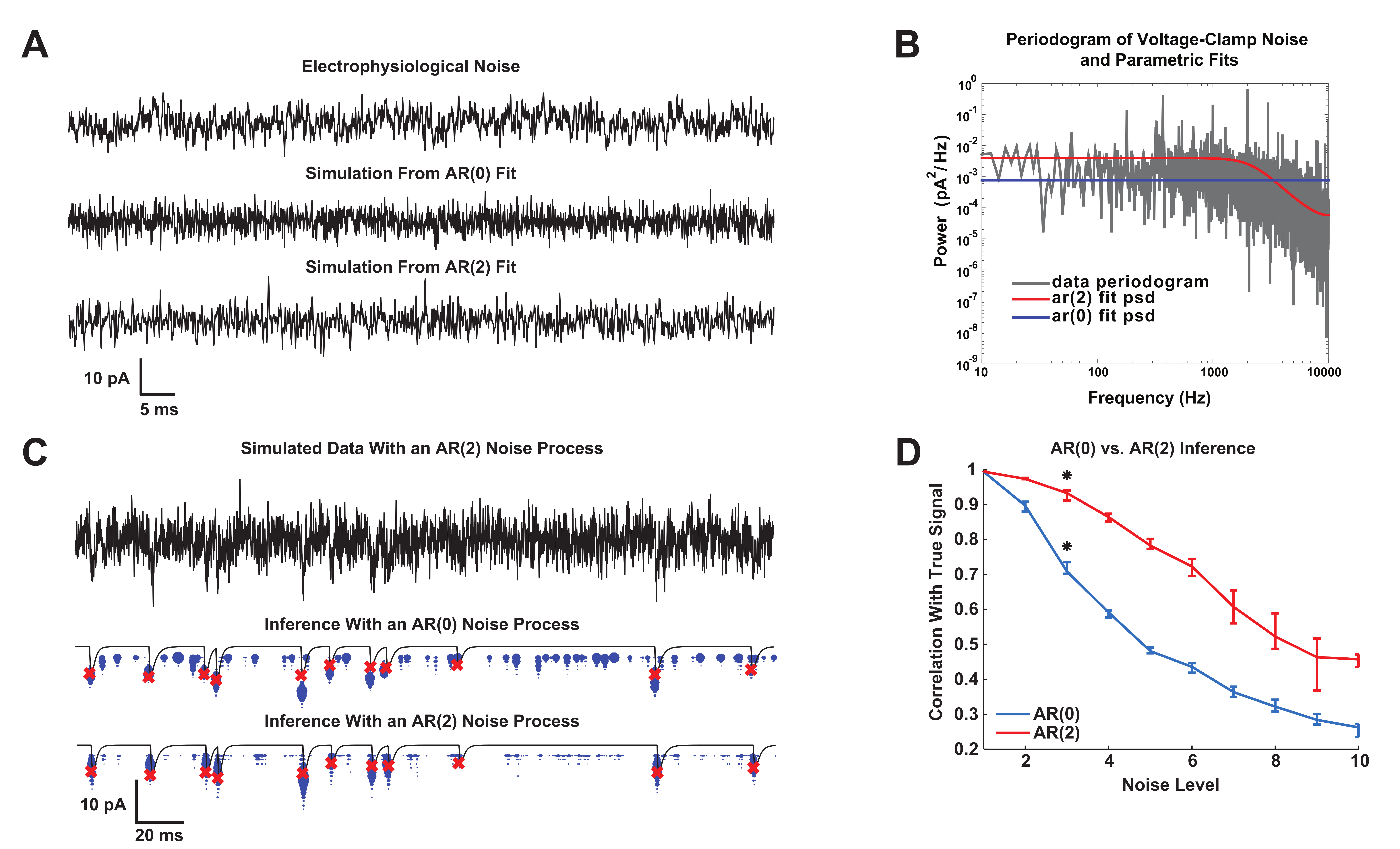}}  
  \end{center}    
  \captionsetup{width=\textwidth}
\captionsetup{font={footnotesize}}
\caption{Validation of the AR($p$) noise model and inference. \textbf{A} Top: a real voltage-clamp recording under ``event-free" conditions such that the recording is dominated by noise. Middle: simulated noise from an AR($0$) fit to the real example. Bottom: simulated noise from an AR($2$) fit to the real example.  The AR($2$) example better captures the structure of real noise.  \textbf{B} Grey: periodogram estimate of the power spectral density (PSD) of an ``event-free" recording. Blue: a parameteric fit to the PSD with an AR($0$) model. Red: a parameteric fit to the PSD with an AR($2$) model. 
\textbf{C} Top: a simulated voltage-clamp recording with AR($2$) noise. Middle, black: The true current for the top trace. Red: The true time-amplitude coordinate for each PSC. Blue: A bivariate histogram reflecting the estimated time-amplitude posterior distribution using an AR($0$) noise model (uses a small, but non-zero minimum event size threshold). Bottom: same as in the middle trace but inference is performed with an AR($2$) noise model. 
\textbf{D} Curves depict accuracy of inference as a function of SNR level (ranging an order of magnitude, with 1 indicating low noise relative to size of events and 10 indicating noise that has marginal variance larger than the signal) for AR(0) vs AR(2) model-based inference. The measure of accuracy is correlation coefficient between true (simulated) trace and estimate of posterior mean.  
The asterisk indicates a biologically realistic SNR level equal to the example in \textbf{C}. For each point on the curve, we simulate timeseries with random event-times and amplitudes (traces are median and inter-quartile range over 10 repeats).  Both algorithms do very well in the high SNR regime.  As soon as noise level begins to make inference difficult, both algorithms begin to lose accuracy.  However the AR($2$) model inference degrades much more slowly.
}
\label{fig:noise_sim}
\end{figure}

We have found the choice of noise model to be critical when analyzing voltage-clamp data.
It is common to use {\it i.i.d} Gaussian noise for neurophysiological time-series (\cite{paninski2012inferring,richardson2008measurement}) , but very often the noise can exhibit temporal correlations.
To obtain a recording of a voltage-clamp noise process which contains no events, we recorded from a neuron exposed to an excitatory synaptic blocker (Kynurenic Acid, 4mM) while holding the cell near the inhibitory reversal potential (-70 mV).
Under these conditions, the recording should be relatively event free; nonetheless the recording shows clear temporal correlations (Figure \ref{fig:noise_sim}\textbf{A}).
In the context of deconvolving these data, it is primarily this correlated noise that drives false positives because it can have similar features to PSCs.

To better capture the structure of voltage-clamp noise, we used the more general AR process which we found was sufficiently flexible and expressive to represent the types of noise we encountered. Specifically, an AR($2$) model balanced model expressiveness and computational cost. Figures \ref{fig:noise_sim}\textbf{A} and \ref{fig:noise_sim}\textbf{B} show that the extra structure in the AR($2$) model does in fact provide a better description of the data. 

For a systematic validation, we performed a comparison between the AR(0) and AR(2) inference for many simulated traces {(10 x 1s traces per noise level)}.  Specifically, we can simulate traces with random event times and with various levels of AR(2) noise added to the traces (i.e., varying SNR).  In these simulations, events are naturalistic in that they have variability in their distribution of amplitudes and time constants, and events may overlap (\ref{fig:noise_sim}\textbf{C}).  For inference with either noise model, performance is similar when there is low levels of noise (or for specific AR(2) noise process parameters that result in only weak noise autocorrelation, not shown), but the inference results diverge dramatically when the noise is larger in magnitude.  To summarize inference, we examine the correlation between the posterior mean trace and the ``true" simulated, noiseless event trace (\ref{fig:noise_sim}\textbf{D}).  
For biologically realistic AR(2) noise structure and magnitude (indicated by an asterisk in \ref{fig:noise_sim}\textbf{D}), inference with the AR(2) model indeed performs considerably better than with the AR(0) model.
Note that in the high-SNR limit, simple methods like template matching algorithms, greedy optimization, or other direct optimization of the model likelihood can perform well enough, so sample-based inference would be computationally excessive.  However, this validation demonstrates that in the biologically realistic noise regime, noise is sufficiently large and structured for proper inference to be useful.  

\subsection{Comparison with other methods on spontaneous and evoked PSCs}

\begin{figure}[!h]
\begin{center}
  \subfloat{\includegraphics[width=1\textwidth]{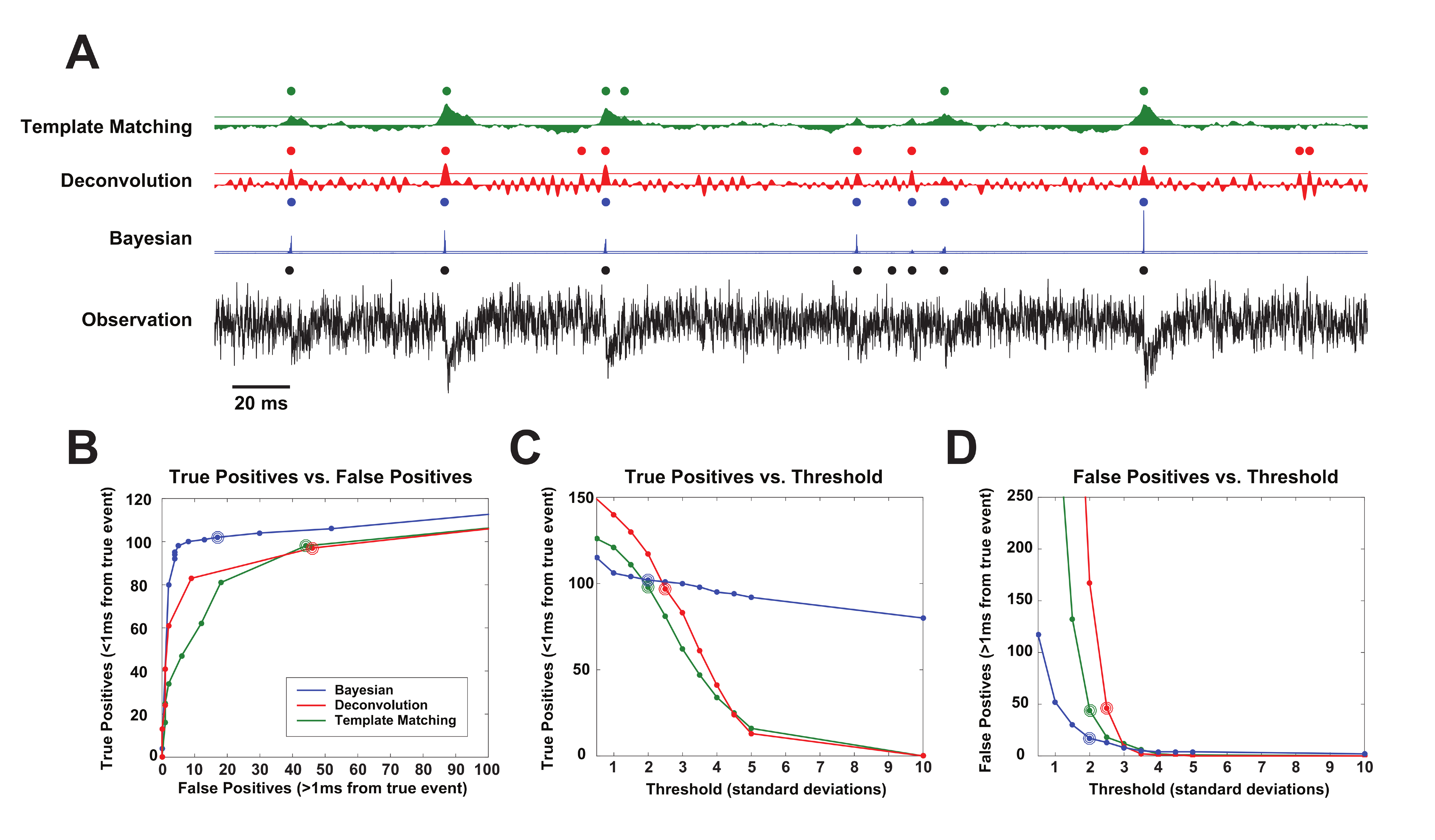}}  
  \end{center}    
  \captionsetup{width=\textwidth}
\captionsetup{font={footnotesize}}
\caption{Comparison of methods on simulated data.  \textbf{A} An example of inference/detection results for each method on a simulated voltage-clamp recording. For each method we show the output time-series for that method (i.e. the score template matching, deconvolved trace for deconvolution, and the posterior of event times for the Bayesian approach), estimated event times, and the threshold used to determine those event times (for the Bayesian trace the threshold is so low that it can't be seen above the baseline posterior). { \textbf{B} Parametric curve showing the number of true positives and false positives as a function of peak detection threshold. The results are from detection across 10 simulated traces with a total of 180 events ranging from 0.5 to 10 pA.} The highlighted point in each line corresponds the threshold used in \textbf{A}. \textbf{C} Same as in \textbf{B} except showing the true positive count as a function of the threshold. \textbf{D} same as in \textbf{C} except showing the false positive count.}
\label{fig:methods_comparison}
\end{figure}

We also compared our method to two common PSC detection algorithms:  a standard template-based approach \cite{clements1997detection} and a deconvolution approach (a Wiener Filter \cite{wiener1949}, which we found tended to improve upon the slightly simpler inverse filter used in \cite{pernia2012deconvolution}).
{Although \cite{clements1997detection} is nearly twenty years old and is often outperformed by deconvolution approaches, it is implemented in two commonly used software packages for the analysis of electrophysiological data: Axograph and pClamp.
Therefore it is still regularly used when PSC detection is performed \cite{Mardinly2016,Medelin2016}.}
When testing these algorithms, we attempted to give each its best chance to perform well.
Specifically, that means that the template-based and deconvolution methods were given the average of the true underlying events as a template. We gave the Wiener Filter the true noise power spectral density for simulated data. For real data with high enough spontaneous rates, it was difficult to find a ``quiet" section of the trace to estimate the noise power spectral density (PSD); therefore we provided an AR fit to the noise from Bayesian inference for the Wiener Filter's noise PSD. 
For simulated data with Bayesian inference, we provided the true priors on $\tau^r$ and $\tau^d$ that were used to generate the data.

First, we simulated a test set of recordings with realistic levels of noise and with relatively low SNR PSCs.  
Each of the three algorithms produces a time series equal in length to the input recording that is something like a score that an event is happening at that point in time. For the template-based method the output represents the goodness-of-fit to the event template at that point in time, and for the deconvolution the output is an estimate of the event amplitude at that point in time. For the Bayesian approach the output time series is the marginal posterior of an event at each sample. 
Importantly, in practice the output for the template-matching and deconvolution methods only have an empirically reasonable interpretation in the high SNR case. 
To obtain estimates of event times for each simulated trace for each method, we { detected peaks above some threshold in} these time series as a function of their standard deviations.
Figure \ref{fig:methods_comparison}\textbf{A} shows an example of results for each algorithm on simulated data.
The threshold for each method's output is shown as a horizontal line, and inferred events are shown as dots. (The threshold for the Bayesian approach is difficult to see as it is very close to zero.) 

By varying the threshold and counting the number of true positives and false positives, we can get a sense of how noisy the output is from each approach.
The Bayesian approach is able to accurately detect more events while accumulating fewer false positives than the other methods (Figure \ref{fig:methods_comparison}\textbf{B}).
{ Importantly, it also maintains similar levels of true and false positives as the threshold is varied (Figures \ref{fig:methods_comparison}\textbf{C} \& \ref{fig:methods_comparison}\textbf{D}), indicating that our approach is more robust to the choice of threshold. This is achieved because of the more precise and less noisy representation of PSC timing produced by the Bayesian method which can be seen in the sharpness of the peaks in the Bayesian trace in Figure \ref{fig:methods_comparison}\textbf{A} as compared to the other methods.
The average error for true positives was also smallest for the Bayesian method (see Supplemental Figure 1).}
While there exists a threshold for the template-matching and deconvolution methods that performs quite well, small deviations from this value lead to vastly more false positives or less true positives (Figures \ref{fig:methods_comparison}\textbf{C} \& \textbf{D}).
When applying these methods to real data, it may not be possible to finely tune the threshold parameter on a per dataset basis since ground truth information is not available. This can be especially troublesome when online or closed-loop analysis is desired.

\begin{figure}[!t]
\begin{center}
  \subfloat{\includegraphics[width=\textwidth]{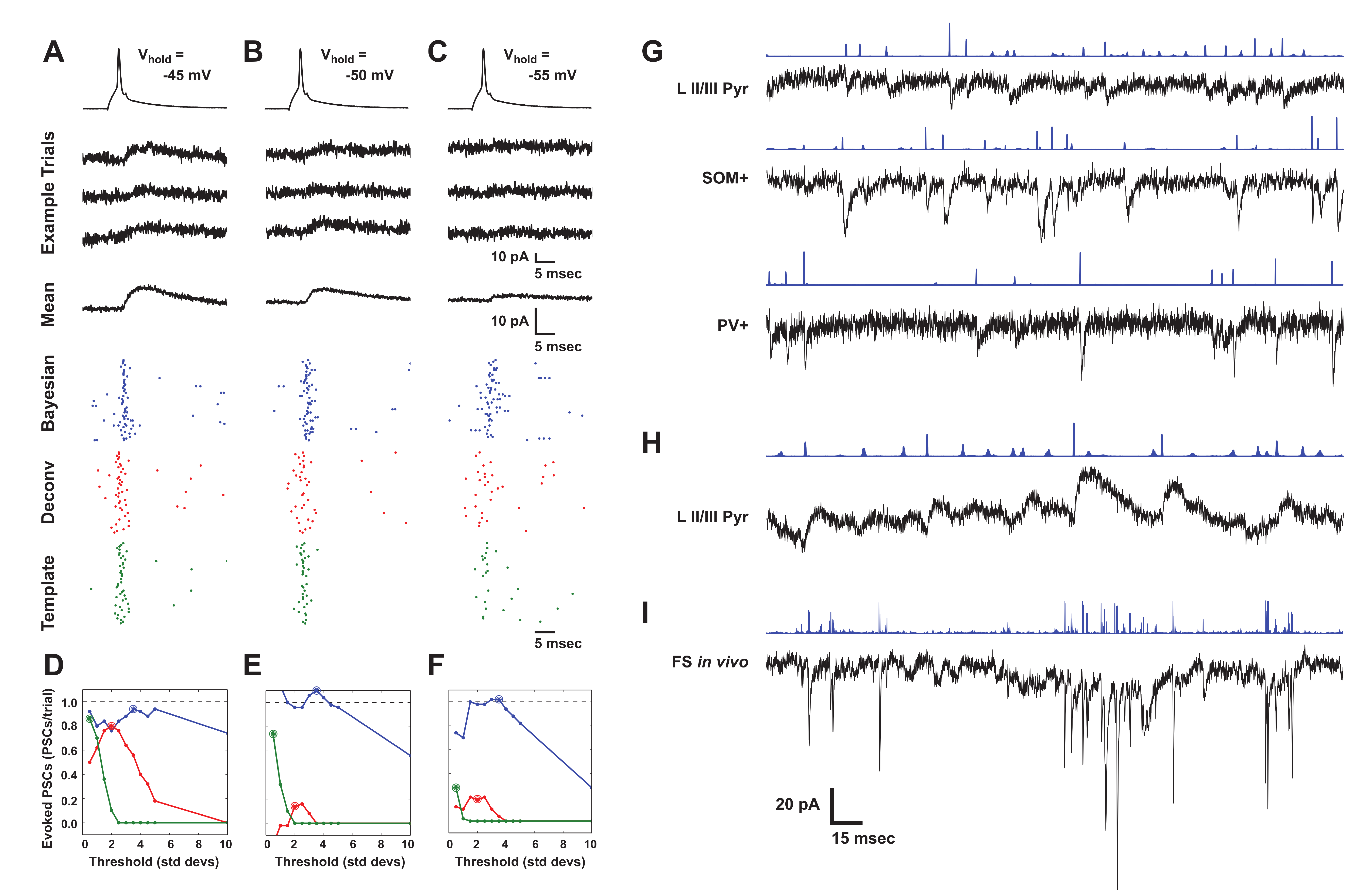}}  
  \end{center}    
  \captionsetup{width=\textwidth}
\captionsetup{font={footnotesize}}
\caption{Results on real voltage-clamp recordings. \textbf{A-F} Comparison to other detection methods on real data while modulating the SNR. \textbf{A} Top: a single trial of the spiking presynaptic cell firing from direct current injection. Middle traces: three example trials of the connected postsynaptic cell responding with an IPSC. The postsynaptic cell is being clamped at -45 mV. Below the examples is the mean trace over all 50 trials. Bottom rasters: estimated events for 50 trials for each detection method. \textbf{B-C} Same as in \textbf{A} except with the postsynaptic cell held at -50 mV (\textbf{B}) and -55 mV (\textbf{C}). The scales for the example traces are all the same across \textbf{A-C} and likewise for the mean traces. \textbf{D-F} The average number of evoked events counted in a window around the spike time (1.0 msec before to 5.0 msec after the spike) above a baseline rate of detected events per trial as a function of threshold for each method. Colors and symbols as in \textbf{2C-B}. The dotted horizontal line represents perfect performance of one extra event detected per trial. \textbf{G-I} Examples of results on several types of real data. \textbf{G} Spontaneous EPSCs detected in a layer II/III pyramidal cell, top, a SOM+ cell, middle, and a PV+ cell, bottom. \textbf{H} Spontaneous IPSCs in a layer II/III pyramidal cell. \textbf{I} Spontaneous EPSCs and IPSCs detected in an \textit{in vivo} recording from a FS cell.}
\label{fig:methods_comparison_real}
\end{figure}

We next compared methods on real voltage-clamp recordings in which we could modulate the SNR of an event physiologically. To achieve this, we made paired patch recordings of a PV+ cell and a layer V pyramidal cell until we found an inhibitory connection from the fast-spiking cell onto the pyramidal cell with a probability of a postsynaptic event extremely close to $1.0$. We then moved the holding potential for the postsynaptic cell towards the reversal potential for the inhibitory current. In this way, we could obtain ground truth data in which we had direct control over the SNR (Figure \ref{fig:methods_comparison_real}\textbf{A}, \textbf{B}, \& \textbf{C}, top traces).

We ran 50 trials each at three holding potentials representing relatively high, medium, and low SNR regimes and detected events using all three methods. Only the Bayesian approach was sensitive enough to detect events reliably in the low SNR case (Figure \ref{fig:methods_comparison_real}\textbf{A}, \textbf{B}, \& \textbf{C}, rasters). Similar to the simulated data, the Bayesian approach was also the only method which was robust to the thresholding parameter, indicating that the posterior over event times is highly peaked. As expected, as the SNR decreases, the ability to accurately detect the timing of the event decreases.

Any approach will have hyperparameters that must be selected, and the Bayesian approach allows for tuning of hyperparameters corresponding to prior distributions on model parameters.  We show that a { single set of prior distribution hyperparameters can perform well across several physiological regimes by running inference on traces from different cell types and under different recording conditions while holding the hyperparameters constant. In \ref{fig:methods_comparison_real}\textbf{G} we show that with a single set of prior parameters our method can detect EPSCs across three different cell types: a layer II/III pyramidal cell, a SOM+ interneuron, and PV+ interneuron. Despite the differing statistics in each of these traces (event features and noise), event detection performs well.
In \ref{fig:methods_comparison_real}\textbf{H} we show results for spontaneous IPSCs in a layer II/III pyramidal cell. For these results, we used the same prior parameters as in \ref{fig:methods_comparison_real}\textbf{G} except that we increased the adjusted the bounds for the time constants to account for the slower kinetcs of IPSCs. Finally, in \ref{fig:methods_comparison_real}\textbf{I} we show results on an FS cell recorded \textit{in vivo}. The prior parameter settings here are similar to those in \ref{fig:methods_comparison_real}\textbf{G} except that we had to increase the rate prior on the number of events. We stress that there is no correct prior setting, but that the priors allow the user to trade off between objectives like true versus false positives (e.g., through the rate parameter or minimum event amplitude). For reference, we include the hyperparameter settings for all inference results in Supplmentary Table 1.}

\subsection{Extension 1: Direct optical stimulation artifact}

For certain experiments, it may be productive to actively drive cells to fire in order to study circuit properties. In particular,  we may want to combine voltage-clamp recordings with spatiotemporally structured optical stimulation of a putative presynaptic population of neurons. 
This can induce optically evoked artifacts from direct optical currents in the voltage-clamp recording. As an example, one could express an optogenetic channel pan-neuronally which would allow the simultaneous mapping EPSCs and IPSCs onto one cell. Under these conditions, the cell under voltage-clamp will also express opsin and respond to any stimulating light. Similarly, in neurotransmitter uncaging mapping experiments, there will be a direct stimulation of the postsynaptic cell at most stimulation sites close to the cell.

\begin{figure}[H]
\begin{center}
  \includegraphics[width=.9\textwidth]{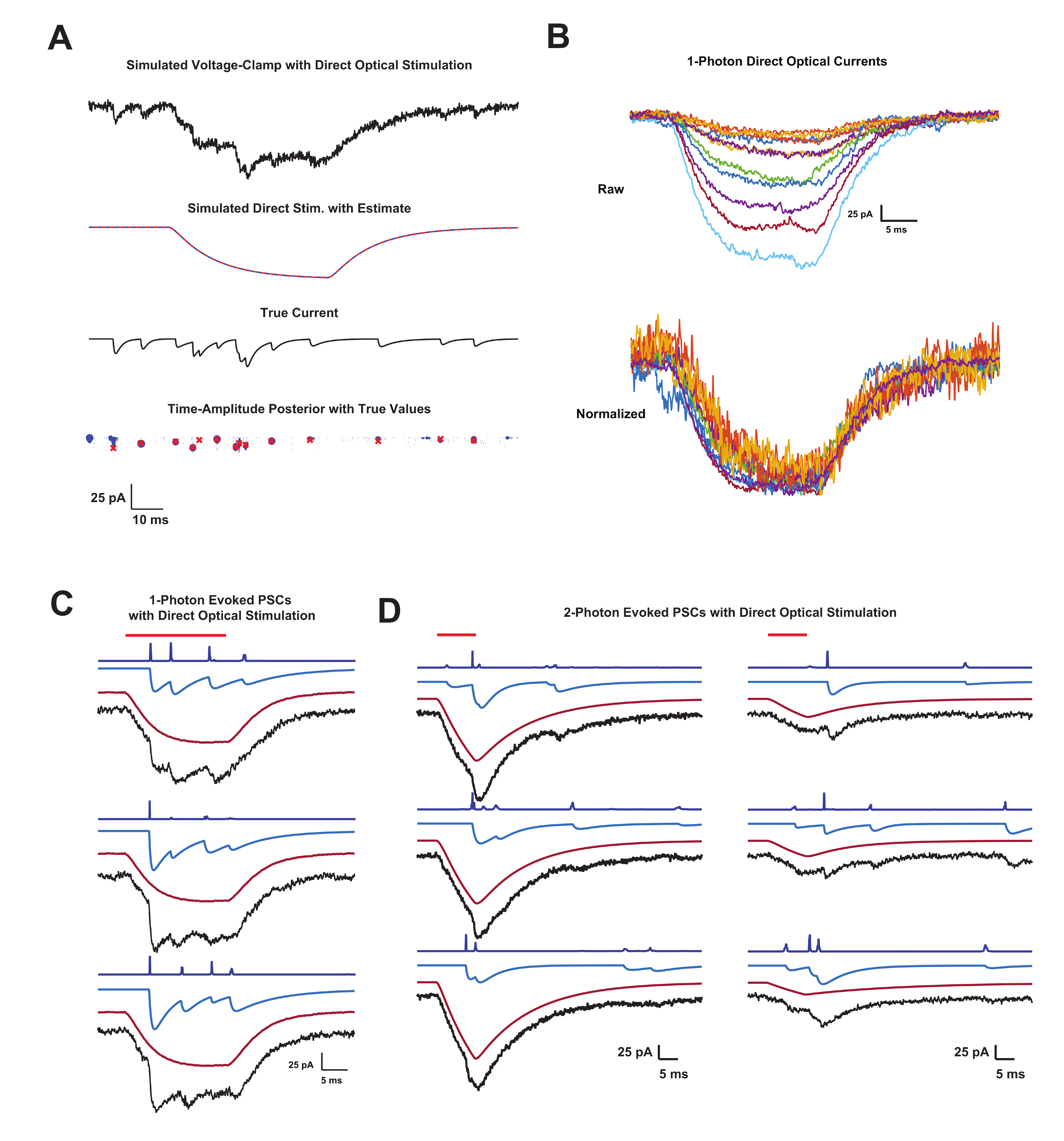}
    \end{center}    
  \captionsetup{width=1\textwidth}
\captionsetup{font={footnotesize}}
\caption{Removing direct stimulation artifacts from active mapping data. \textbf{A} Top: a simulated voltage-clamp recording that is contaminated by a large direct, optical current. Top middle: the true direct current (blue) with an estimate of the inferred direct current (dashed, red). Bottom middle: the true current used in the top trace. Bottom, the inferred amplitude-time posterior (blue) with true amplitude-time coordinates overlaid (red X's). \textbf{B} Top: direct optical currents evoked in a single cell by stimulating different locations with one-photon excitation and a DMD. Bottom: same as above but each trace is normalized to have the same amplitude. \textbf{C-D} PSC detection with direct optical stimulation on real mapping data. \textbf{C} Inference results for one-photon, DMD-based mapping data with direct stimulation contamination. All three trials are for a single stimulation site. Red line shows when the stimulating laser is on. Dark blue trace shows the posterior over event times. Light Blue trace shows the MAP estimate for the synaptic currents. Maroon trace, MAP estimate for the direct stimulation. Black, raw voltage-clamp observation. \textbf{D} Inference results for two-photon, SLM-based mapping data with direct stimulation contamination. Left, three trials from a particular stimulation location which shows putative evoked EPSCs riding on top of direct stimulation with inference.  Right, same as the results to the left but for a stimulation location further from the cell. }
\label{fig:direct_stim_real}
\end{figure}

We show that our approach is able to decompose a simulated trace with a direct optical current in Figure \ref{fig:direct_stim_real}.
Specifically, we simulated a trace consisting of many EPSCs with an additive direct optical stimulation current, consistent with eqn. \ref{artifact_eqn}.  On this simulated data, inference performs very well in terms of extracting events simultaneously with artifact isolation.

In Figure \ref{fig:direct_stim_real}\textbf{B}-\textbf{D}, we present results on real data obtained by combining voltage-clamp recording with one-photon stimulation via a digital micromirror device (DMD) as well as two-photon, holographic stimulation of neurons. In \ref{fig:direct_stim_real}\textbf{B}, we show direct currents at many different stimulation locations with one-photon excitation via a DMD. Under these conditions, the shape of the current is not well characterized by a parameterized template function so we must use an empirically derived template. When normalized, all of the currents have roughly similar shapes, validating our approximation of the artifact as a scaled template. In \ref{fig:direct_stim_real}\textbf{C} we show PSC inference on three trials from a single stimulation location that has putative evoked PSCs overlapping the direct stimulation artifact.
In Figure \ref{fig:direct_stim_real}\textbf{D}, we show similar results except the stimulation is performed using two-photon excitation with a spatial-light modulator (SLM). We found that under these conditions, we were able to use the model in eqn. \ref{artifact_w_shape_eqn} and fit two time constants to the optical current. This allowed for a more flexible fit that could account for the differences in the direct current shape as the stimulation location varied. The two sets of traces and results in \ref{fig:direct_stim_real}\textbf{D} are for three trials at two different locations. These locations were chosen because they contained putative evoked EPSCs.

\subsection{Extension 2: Passive mapping experiment (simulation)}
\label{section_passive_mapping}

\begin{figure}[H]
\begin{center} 
  \subfloat{\includegraphics[width=\textwidth]{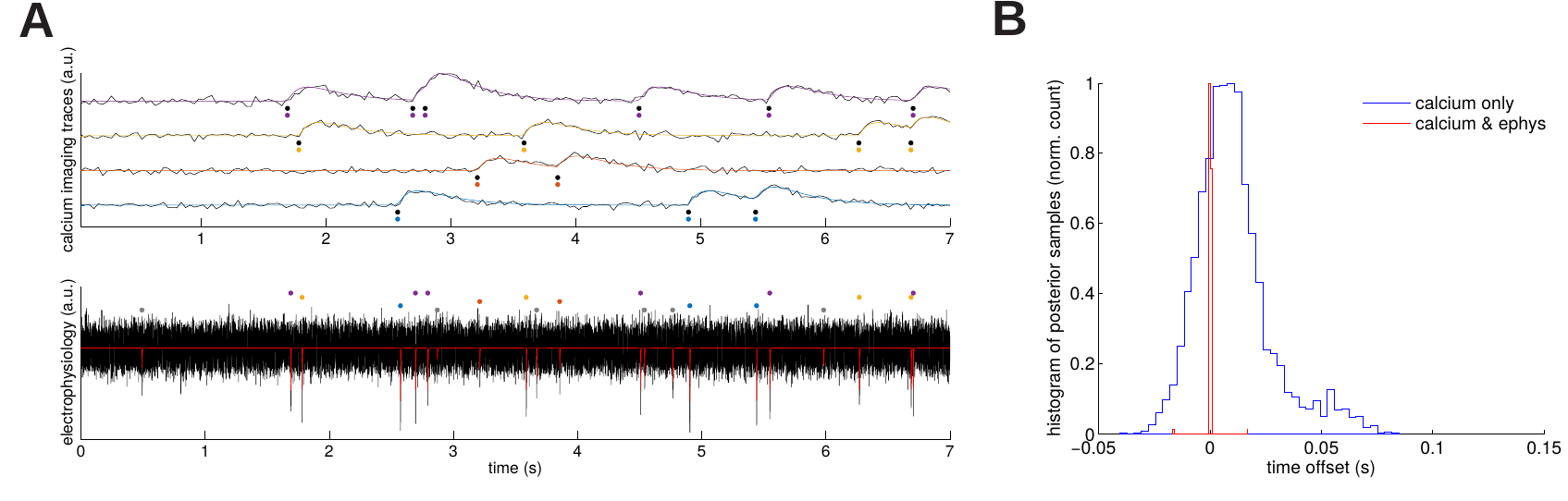}} 
  \end{center}    
  \captionsetup{width=1\textwidth}
\captionsetup{font={footnotesize}}
\caption{Joint inference over simulated calcium and voltage data.  \textbf{A} Observed data and true/inferred event times for four simultaneously observed calcium traces (top) and (for simplicity) just a single voltage trace (bottom).  Colored dots indicate to which trace the algorithm matched each event; gray dots in bottom correspond to voltage events that were detected but (correctly) not matched to any detected calcium events.  In this example the algorithm correctly matched all calcium-observed events.  \textbf{B} Histograms of temporal error in estimated event times.  Given calcium only, the variance of inferred event times is large; incorporating voltage information drastically reduces this variance, indicating significant potential gains in temporal resolution from this Bayesian data fusion approach.}
\label{fig:mapping}
\end{figure}

In addition to electrical recordings from a post-synaptic cell, we may also have calcium indicator available in pre-synaptic cells \cite{aaron2006reverse}.  This allows for a passive mapping experiment from many pre-synaptic cells (optically imaged) to a single post-synaptic cell (patched).   Here we provide an example of the usage of the joint inference procedure (Figure \ref{fig:mapping}\textbf{A}).  We have simulated a population of presynaptic cells which are observed via calcium imaging (SNR for these cells is plausible for a well-tuned setting, and bin size is 35ms).  These cells drive post-synaptic events in the simulated patched cell (noise is biologically plausible, and bin size is .05ms).  In \ref{fig:mapping}\textbf{A}, 4 of 6 candidate presynaptic neurons are observed.  Postsynaptic events evoked by unobserved neurons are inferred on a per event basis whereas events co-occuring with presynaptic events arise from a single variable (see model eqns. \ref{mapping_model}).    

We see that we can recover events jointly from the calcium imaging and electrophysiology, with event identity successfully linked across the two modalities.  In addition, the histogram in the right panel of Figure \ref{fig:mapping}\textbf{B} indicates that combining electrophysiology and calcium imaging tends to yield increased temporal precision of event times compared to inference from calcium imaging alone (this intuitively follows from the fact that the electrophysiology has much higher temporal resolution and the Bayesian inference can combine information across modalities).  This simulation serves as a proof-of-concept that this probabilistic approach can provide meaningful automatic analysis for passive mapping experiments.  

\section{Discussion}

In this work we have presented a probabilistic formulation of the event detection problem for electrophysiological recordings with an emphasis on PSC detection.  This method works on simulated and real data and its performance compares favorably relative to existing methods. We have also shown how the probabilistic model can be incorporated into two extensions applicable for mapping experiments which make use either of optical stimulation or multi-modal fusion of electrophysiology with calcium imaging.

{The PSC detection problem explored in this paper is similar to the more general set of problems in neuroscience (and signal processing more broadly) having to do with inferring events in noisy timeseries.  In neuroscience, recent work has proposed probabilistic, MCMC-based tools for analysis of calcium imaging data \cite{vogelstein2009spike,pnevmatikakis2013bayesian}.  However, we are not aware of similar tools yet being leveraged for analysis of electrophysiology data.  The models developed in this paper are formally similar to those used in the calcium imaging setting, with appropriate modifications to allow for structured AR noise and distinct per event kinetics, as well as detailing the extensions related to the mapping setting.

While there are other discriminative methods that have been proposed for deconvolution (e.g. see \cite{Theis2016} for alternative event-detection methods for calcium imaging), these approaches lack the clear probabilistic generative semantics which facilitate modularity and extension to hierarchical models such as those useful for mapping experiments. 
Our approach also complements related work inferring synaptic inputs in a probabilistic fashion using particle filtering \cite{paninski2012inferring} from voltage traces. The present approach focuses on current traces, allows for per event kinetics, is perhaps simpler to implement, and does not require temporal discretization.  

The inference approach presented in this work makes most sense when PSC rates are relatively low such that overlap is limited.  For very high levels of overlap (i.e. the high-rate case), we do not expect there is always hope to resolve precise timing of single events, and it is likely that approaches more like \cite{vogelstein2010fast} or \cite{paninski2012inferring} may be appropriate.  That said, we have found that our method can still give sensible results even when event rates are high (for example, see Figure \ref{fig:methods_comparison_real}I).}  

Aside from the extensions we considered, other sophisticated prior structure can be incorporated by making the model hierarchical.  For example, in the single trace setting we might expect the events to cluster by pre-synaptic cell identity or cell type.  Even without the additional observed traces employed in our mapping model, it would be conceivable to specify and infer latent source clusters based on structure in the distribution of shape or amplitude of post-synaptic events, depending on what were of most interest scientifically.  In the simplest case, this would result in a mixture model for the events, with each event associated with a latent pre-synaptic source.  We plan to pursue these model extensions in future work.

\section*{Acknowledgements}
Funding for this research was provided by Global Brain Research Award 325398 from the Simons Foundation, ARO MURI W911NF-12-1-0594, and DARPA N66001-15-C-4032 (SIMPLEX); in addition, this work was supported by the Intelligence Advanced Research Projects Activity (IARPA) via Department of Interior/ Interior Business Center (DoI/IBC) contract number D16PC00003.  The U.S. Government is authorized to reproduce and distribute reprints for Governmental purposes notwithstanding any copyright annotation thereon. Disclaimer: The views and conclusions contained herein are those of the authors and should not be interpreted as necessarily representing the official policies or endorsements, either expressed or implied, of IARPA, DoI/IBC, or the U.S. Government. B.S. is supported by a Fannie and John Hertz Foundation Fellowship and an NSF Graduate Research Fellowship.

\section{References}

\bibliography{references.bib}

\begin{thebibliography}{10}
\expandafter\ifx\csname url\endcsname\relax
  \def\url#1{\texttt{#1}}\fi
\expandafter\ifx\csname urlprefix\endcsname\relax\def\urlprefix{URL }\fi
\expandafter\ifx\csname href\endcsname\relax
  \def\href#1#2{#2} \def\path#1{#1}\fi

\bibitem{Rickgauer2014}
J.~P. Rickgauer, K.~Deisseroth, D.~W. Tank, Simultaneous cellular-resolution
  optical perturbation and imaging of place cell firing fields, Nature
  Neuroscience 17~(12) (2014) 1816--1824.

\bibitem{scanziani2009electrophysiology}
M.~Scanziani, M.~H{\"a}usser, Electrophysiology in the age of light, Nature
  461~(7266) (2009) 930--939.

\bibitem{baryehuda1127}
D.~Bar-Yehuda, A.~Korngreen, Space-clamp problems when voltage clamping neurons
  expressing voltage-gated conductances, Journal of Neurophysiology 99~(3)
  (2008) 1127--1136.

\bibitem{shababo2013bayesian}
B.~Shababo, B.~Paige, A.~Pakman, L.~Paninski, Bayesian inference and online
  experimental design for mapping neural microcircuits, Advances in Neural
  Information Processing Systems (NIPS) (2013) 1304--1312.

\bibitem{pnevmatikakis2013bayesian}
E.~A. Pnevmatikakis, J.~Merel, A.~Pakman, L.~Paninski, Bayesian spike inference
  from calcium imaging data, in: Asilomar Conference on Signals, Systems and
  Computers, IEEE, 2013, pp. 349--353.

\bibitem{aaron2006reverse}
G.~Aaron, R.~Yuste, Reverse optical probing (roping) of neocortical circuits,
  Synapse 60~(6) (2006) 437.

\bibitem{jonas1993quantal}
P.~Jonas, G.~Major, B.~Sakmann, Quantal components of unitary {EPSCs} at the
  mossy fibre synapse on {CA3} pyramidal cells of rat hippocampus., The Journal
  of Physiology 472~(1) (1993) 615--663.

\bibitem{ankri1994automatic}
N.~Ankri, P.~Legendre, D.~Faber, H.~Korn, Automatic detection of spontaneous
  synaptic responses in central neurons, Journal of Neuroscience Methods 52~(1)
  (1994) 87--100.

\bibitem{hwang1999automatic}
T.~N. Hwang, D.~R. Copenhagen, Automatic detection, characterization, and
  discrimination of kinetically distinct spontaneous synaptic events, Journal
  of Neuroscience Methods 92~(1) (1999) 65--73.

\bibitem{kudoh2002simple}
S.~N. Kudoh, T.~Taguchi, A simple exploratory algorithm for the accurate and
  fast detection of spontaneous synaptic events, Biosensors and Bioelectronics
  17~(9) (2002) 773--782.

\bibitem{clements1997detection}
J.~Clements, J.~Bekkers, Detection of spontaneous synaptic events with an
  optimally scaled template., Biophysical Journal 73~(1) (1997) 220.

\bibitem{pernia2012deconvolution}
A.~J. Pern{\'\i}a-Andrade, S.~P. Goswami, Y.~Stickler, U.~Fr{\"o}be,
  A.~Schl{\"o}gl, P.~Jonas, A deconvolution-based method with high sensitivity
  and temporal resolution for detection of spontaneous synaptic currents in
  vitro and in vivo, Biophysical Journal 103~(7) (2012) 1429--1439.

\bibitem{guzman2014stimfit}
S.~J. Guzman, A.~Schl{\"o}gl, C.~Schmidt-Hieber, Stimfit: quantifying
  electrophysiological data with python, Frontiers in Neuroinformatics 8~(16)
  (2014) \textit{online}.

\bibitem{richardson2008measurement}
M.~J. Richardson, G.~Silberberg, Measurement and analysis of postsynaptic
  potentials using a novel voltage-deconvolution method, Journal of
  Neurophysiology 99~(2) (2008) 1020--1031.

\bibitem{li2007weighted}
G.-H. Li, M.~F. Jackson, J.~F. MacDonald, Weighted least squares fitting with
  multiple templates for detection of small spontaneous signals, Journal of
  Neuroscience Methods 164~(1) (2007) 139--148.

\bibitem{shi2010novel}
Y.~Shi, Z.~Nenadic, X.~Xu, Novel use of matched filtering for synaptic event
  detection and extraction, PLoS One 5~(11) (2010) e15517.

\bibitem{chib1994bayes}
S.~Chib, E.~Greenberg, Bayes inference in regression models with arma (p, q)
  errors, Journal of Econometrics 64~(1) (1994) 183--206.

\bibitem{wiener1949}
N.~Wiener, Extrapolation, interpolation, and smoothing of stationary time
  series: with engineering applications, Technology press books in science and
  engineering, Technology Press of the Massachusetts Institute of Technology,
  1949.

\bibitem{moller2004}
J.~Moller, R.~P. Waagepetersen, Statistical inference and simulation for
  spatial point processes, CRC Press, 2004.

\bibitem{tan2008}
V.~Tan, V.~Goyal, Estimating signals with finite rate of innovation from noisy
  samples: A stochastic algorithm, IEEE Transactions on Signal Processing
  56~(10) (2008) 5135--5146.

\bibitem{lehmann1998theory}
E.~L. Lehmann, G.~Casella, Theory of point estimation, Vol.~31, Springer
  Science \& Business Media, 1998.

\bibitem{fenno2011}
L.~Fenno, O.~Yizhar, K.~Deisseroth, The development and application of
  optogenetics, Annual Review of Neuroscience 34~(1) (2011) 389--412.

\bibitem{callaway2002}
E.~M. Callaway, R.~Yuste, Stimulating neurons with light, Current Opinion in
  Neurobiology 12~(5) (2002) 587 -- 592.

\bibitem{Shepherd2005}
G.~M.~G. Shepherd, A.~Stepanyants, I.~Bureau, D.~Chklovskii, K.~Svoboda,
  Geometric and functional organization of cortical circuits, Nature
  Neuroscience 8~(6) (2005) 782--790.

\bibitem{Katzel2011}
D.~Katzel, B.~V. Zemelman, C.~Buetfering, M.~Wolfel, G.~Miesenbock, The
  columnar and laminar organization of inhibitory connections to neocortical
  excitatory cells, Nature Neuroscience 14~(1) (2011) 100--107.

\bibitem{mena2015}
G.~Mena, L.~Grosberg, F.~Kellison-Linn, E.~Chichilnisky, L.~Paninski,
  Large-scale multi electrode array spike sorting algorithm introducing
  concurrent recording and stimulation, in: NIPS workshop on Statistical
  Methods for Understanding Neural Systems, 2015.

\bibitem{vogelstein2009spike}
J.~T. Vogelstein, B.~O. Watson, A.~M. Packer, R.~Yuste, B.~Jedynak,
  L.~Paninski, Spike inference from calcium imaging using sequential monte
  carlo methods, Biophysical Journal 97~(2) (2009) 636--655.

\bibitem{neal1993probabilistic}
R.~M. Neal, Probabilistic inference using Markov chain Monte Carlo methods,
  Department of Computer Science, University of Toronto Toronto, CA, 1993.

\bibitem{gelman2014bayesian}
A.~Gelman, J.~B. Carlin, H.~S. Stern, D.~B. Rubin, Bayesian data analysis,
  Vol.~2, Taylor \& Francis, 2014.

\bibitem{Lay2012}
D.~C. Lay, Linear Algebra and Its Applications, 4th Edition, Pearson, 2012.

\bibitem{rosenthal2011optimal}
J.~S. Rosenthal, Optimal proposal distributions and adaptive mcmc, Handbook of
  Markov Chain Monte Carlo (2011) 93--112.

\bibitem{Pluta2015}
S.~Pluta, A.~Naka, J.~Veit, G.~Telian, L.~Yao, R.~Hakim, D.~Taylor, H.~Adesnik,
  A direct translaminar inhibitory circuit tunes cortical output, Nature
  Neuroscience 18~(11) (2015) 1631--1640.

\bibitem{paninski2012inferring}
L.~Paninski, M.~Vidne, B.~DePasquale, D.~G. Ferreira, Inferring synaptic inputs
  given a noisy voltage trace via sequential {Monte Carlo} methods, Journal of
  Computational Neuroscience 33~(1) (2012) 1--19.

\bibitem{Mardinly2016}
A.~R. Mardinly, I.~Spiegel, A.~Patrizi, E.~Centofante, J.~E. Bazinet, C.~P.
  Tzeng, C.~Mandel-Brehm, D.~A. Harmin, H.~Adesnik, M.~Fagiolini, M.~E.
  Greenberg, Sensory experience regulates cortical inhibition by inducing igf1
  in vip neurons, Nature 531~(7594) (2016) 371--375.

\bibitem{Medelin2016}
M.~Medelin, V.~Rancic, G.~Cellot, J.~Laishram, P.~Veeraraghavan, C.~Rossi,
  L.~Muzio, L.~Sivilotti, L.~Ballerini, Altered development in gaba co-release
  shapes glycinergic synaptic currents in cultured spinal slices of the
  {SOD1G93A} mouse model of als, The Journal of Physiology (2016).

\bibitem{Theis2016}
L.~Theis, P.~Berens, E.~Froudarakis, J.~Reimer, M.~Roman-Roson, T.~Baden,
  T.~Euler, A.~S. Tolias, M.~Bethge, Benchmarking spike rate inference in
  population calcium imaging, Neuron 90~(3) (2016) 471--482.

\bibitem{vogelstein2010fast}
J.~T. Vogelstein, A.~M. Packer, T.~A. Machado, T.~Sippy, B.~Babadi, R.~Yuste,
  L.~Paninski, Fast nonnegative deconvolution for spike train inference from
  population calcium imaging, Journal of Neurophysiology 104~(6) (2010)
  3691--3704.

\end{thebibliography}

\section{Supplemental Table 1: Inference Hyperparameters}
\begin{center}
    \begin{tabular}{| l | p{1cm} | p{1cm} | p{1cm} | p{1cm} | p{1cm} | p{1cm} | p{2cm} |p{1cm} |}
    \hline
    &$a_{min}$ (pA) & $a_{max}$ (pA) & $\tau_{min}^r$ (msec) & $\tau_{max}^r$ (msec) & $\tau_{min}^d$ (msec) & $\tau_{max}^d$ (msec) & $\lambda_n$ events/sec & \# Gibbs Sweeps \\ \hline
    Figure 2C-D  & 1.5 & Inf & 0.25 & 2.50 & 0.25 & 2.50 & 0.0013 & 1000 \\ \hline
    Figure 3 & 0.01 & Inf & 0.05 & 1.00 & 0.50 & 10.00 & 2.00 & 1000 \\ \hline
    Figure 4A-C & -Inf & Inf & 0.25 & 3.00 & 1.00 & 30.00 & 0.02 & 2000 \\ \hline
	Figure 4G LII/III Pyr & 0.5 & Inf & 0.25 & 1.50 & 1.00 & 5.00 & 2.00 & 2000 \\ \hline
	Figure 4G SOM+ & 0.5 & Inf & 0.25 & 1.50 & 1.00 & 5.00 & 2.00 & 2000 \\ \hline
	Figure 4G PV+ & 0.5 & Inf & 0.25 & 1.50 & 1.00 & 5.00 & 2.00 & 2000 \\ \hline
	Figure 4H & 0.5 & Inf & 1.00 & 3.00 & 5.00 & 30.00 & 2.00 & 2000 \\ \hline
	Figure 4I & -Inf & Inf & 0.05 & 1.00 & 5.00 & 7.50 & 2000 & 2000 \\ \hline
	Figure 5A & 5.0 & Inf & 0.05 & 0.50 & 0.50 & 10.00 & 200 & 2000 \\ \hline
	Figure 5C & 5.0 & Inf & 0.05 & 0.50 & 0.50 & 10.00 & 0.20 & 2000 \\ \hline
	Figure 5D & 2.5 & Inf & 0.10 & 1.00 & 1.00 & 8.00 & 0.20 & 3000 \\ \hline
    \end{tabular}
\end{center}

\pagebreak

\section{Supplemental Figure 1: Timing Error Distributions}

\begin{figure}[h!]
\begin{center}
  \subfloat{\includegraphics[width=1\textwidth]{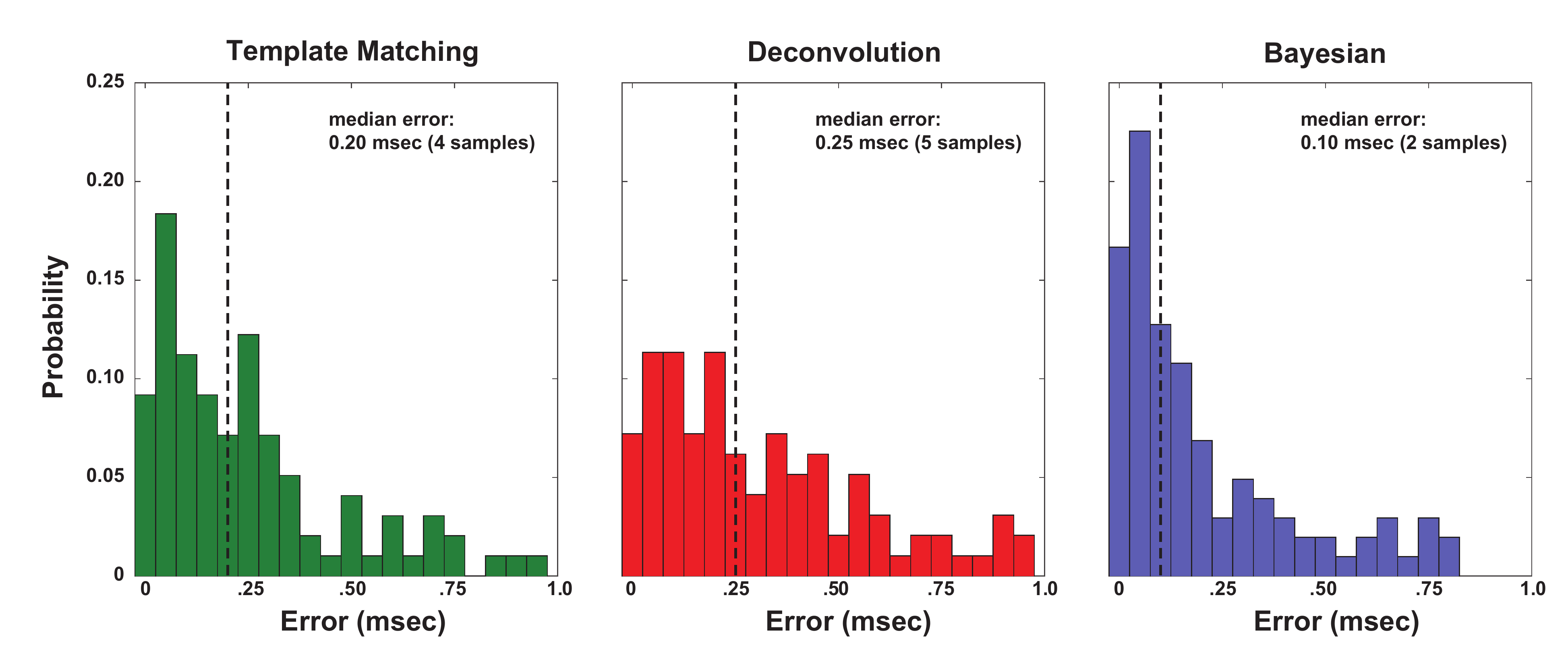}}  
  \end{center}    
  \captionsetup{width=1\textwidth}
\captionsetup{font={footnotesize}}
\caption{The distributions of timing errors from the results corresponding to the highlighted points in Figures 3B-D. The histograms have been normalized such that the areas for each histogram sum to one. The vertical dashed line shows the median value for each distribution. Although the Bayesian method models the timing of events in continuous time, the resolution of detection for all methods was set to the sampling resolution of the data (20 kHz) as a result of the peak detection process.}
\label{fig:temporal_errs}
\end{figure}

\end{document}